# Testing of Front-End Readout Prototype ASICs Designed for WCDA in LHAASO

Shengzhi Zhou, Lei Zhao, *Member, IEEE*, Yuxiang Guo, Ruoshi Dong, Boyu Cheng, Jiajun Qin, Sifan Qian, Shubin Liu, *Member, IEEE*, Qi An, *Member, IEEE*

*Abstract*—The water Cherenkov detector array (WCDA) is one of the key detectors in the large high altitude air shower observatory (LHAASO), which is proposed for very high gamma ray source survey. In WCDA, there are more than 3000 photomultiplier tubes (PMTs) scattered under water in an area of 80000 m$^2$. As for the WCDA readout electronics, both high precision time and charge measurement is required over a large dynamic range from 1 photon electron (P.E.) to 4000 P.E. To reduce the electronics complexity and improve the system reliability, a readout scheme based on application specific integrated circuits (ASICs) is proposed. Two prototype ASICs were designed and tested. The first ASIC integrates amplification and shaping circuits for charge measurement and discrimination circuits used for time measurement. The shaped signal is further fed to the second ADC ASIC, while the output signal from the discriminator is digitized by the FPGA-based time-to-digital converter (TDC). Test results indicate that time resolution is better than 250 ps RMS, and the charge resolution is better than 10% at 1 P.E., and 1% at 4000 P.E. which meets the requirements of the LHAASO WCDA.

## I. Introduction

THE large high altitude air shower observatory (LHAASO) aims to discover the cosmic ray origin and perform the advanced scientific research in the field of high-energy physics and astronomy. The water Cherenkov detector array (WCDA) is one of the key detectors in the LHAASO, and it is composed of three water ponds with more than 3000 photomultiplier tubes (PMTs). Due to the research demands, both high precision time and charge measurements are required over a large dynamic range from 1 Photon Electron (P.E.) to 4000 P.E. The detectors in the LHAASO are shown in Fig. 1, including electromagnetic particle detector (ED), muon detector (MD), wide field of view Cherenkov telescopes array (WFCTA) and WCDA [1-7].

Manuscript is received on February 2, 2018. This work was supported by the National Natural Science Foundation of China (11722545) and the CAS Center for Excellence in Particle Physics (CCEPP).

The authors are with the State Key Laboratory of Particle Detection and Electronics, University of Science and Technology of China, Hefei 230026, China, and Department of Modern Physics, University of Science and Technology of China, Hefei 230026, China (telephone: 086-0551-63607746, corresponding author: Lei Zhao, e-mail: zlei@ustc.edu.cn).

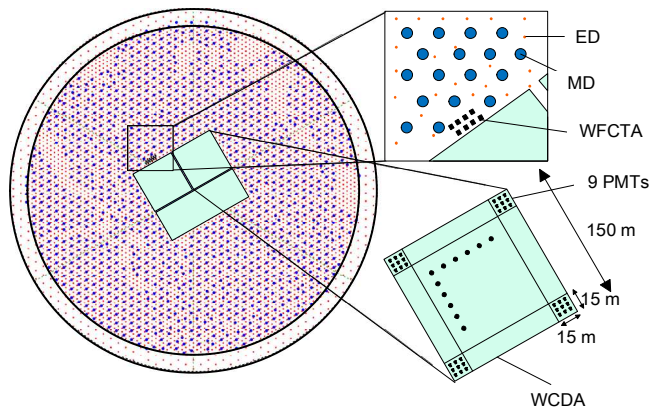

Fig. 1. Detectors in the LHAASO.

The rest of the paper is organized as follows. In Section II, ASIC-based front-end readout electronics is introduced and the amplification and shaping ASIC is presented in Section III. In Section IV, the SAR ADC ASIC is explained in detail. The results of the combined tests of two ASICs are provided in Section V. Lastly, a brief conclusion is given in Section VI.

## II. ASIC-Based Front-End Readout Electronics

The readout system of the LHAASO WCDA is shown in Fig. 2. The output PMT signals are fed to the front-end electronics, and then the digitalized data are sent to the data acquisition system (DAQ) through switches. One front-end electronics module can deal with up to nine PMT output signals [1, 2, 8].

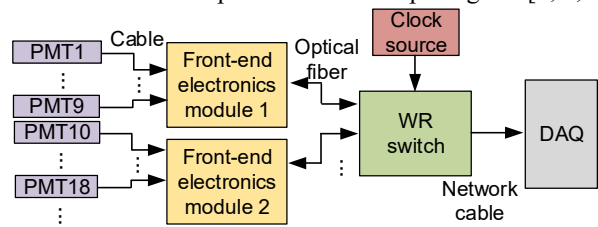

Fig. 2. LHAASO WCDA readout system.

To simplify the circuit structure of the readout electronics, a readout scheme based on application specific integrated circuit (ASIC) is proposed. After the investigation of different ASICs used in large-scale physics experiments all over the world, including switched capacitor array (SCA) ASIC [9-11], time-over-threshold (TOT) ASIC [12-14] and the amplification and shaping ASIC [15-17], one technical route based on amplification and shaping, analog-to-digital conversion combined with digital peak detection is proposed [2]. The architecture of the ASIC-based PMT front-end readout electronics is shown in Fig. 3. Two prototype ASICs that integrate the amplification and shaping circuits and analog-to-digital converter (ADC) circuits are designed respectively. Charge measurement is performed by using the combination of the anode and dynode channels to cover the dynamic range from 1 to 4000 P.E. in the amplification and shaping ASIC. The shaper output signals are then digitized by the second ASIC, which is a successive approximation (SAR) ADC [18]. The digitized amplitude information is transformed to the field programmable gate array (FPGA) for peak detection. As for the time measurement, the PMT anode signal is fed to the discrimination circuits inside the amplification and shaping ASIC, and then digitized by the FPGA-based time-to-digital converter (TDC).

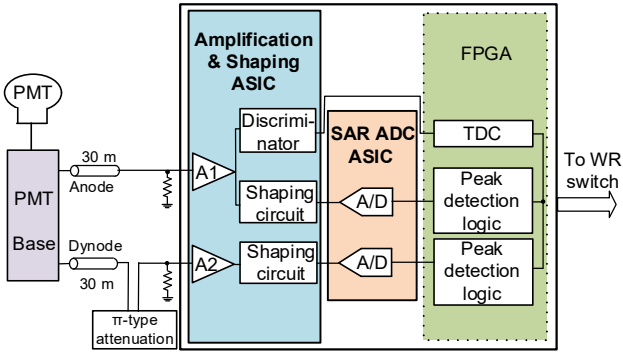

Fig. 3. Architecture of the ASIC-based PMT front-end readout electronics.

## III. AMPLIFICATION AND SHAPING ASIC

### A. Structure of Amplification and Shaping ASIC

The amplification and shaping ASIC is designed with three anode channels and three dynode channels based on the Global Foundry 350-nm CMOS dual gate technology.

The structure of the amplification and shaping ASIC is shown in Fig. 4, wherein it can be seen that it contains the preamplifiers, RC$^4$ shaping circuits, output buffers, and discriminators. In the charge measurement, two channels are used to read out the signals from the anode and dynode of one PMT to achieve a large dynamic range. On the other hand in the time measurement, both low and high thresholds are used, whereas the latter is used to avoid the deterioration of time resolution around the baseline of a large input signal [8].

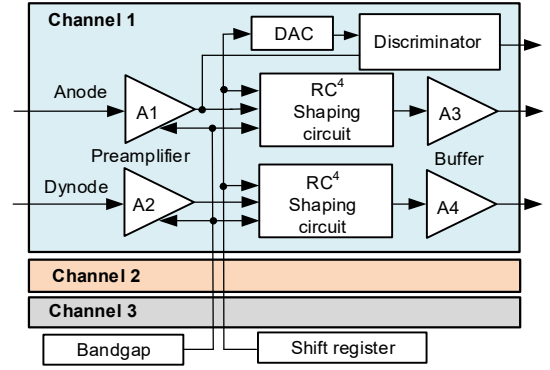

Fig. 4. Structure of the amplification and shaping ASIC.

### B. Anode Channel and RC Node Simulation Waveform

The structure of the anode channel is shown in Fig. 5. The preamplifier is a voltage amplifier because this kind of amplifiers has the high input impedance and wide bandwidth, which enables achievement of a high-precision time measurement. In the filter circuit, RC$^4$ shaping circuit is used because the signal-to-noise ratio (SNR) of the output signal is higher than the SNR of the Sallen Key shaping circuit. Also, the time constant can be adjusted to 20 ns, 30 ns, 40 ns and 50 ns. Considering the large power consumption caused by the large load, the output buffer uses the class-AB structure to reduce the static power consumption. The output signal from the preamplifier is also fanned out for the discriminators of high threshold and low threshold [18].

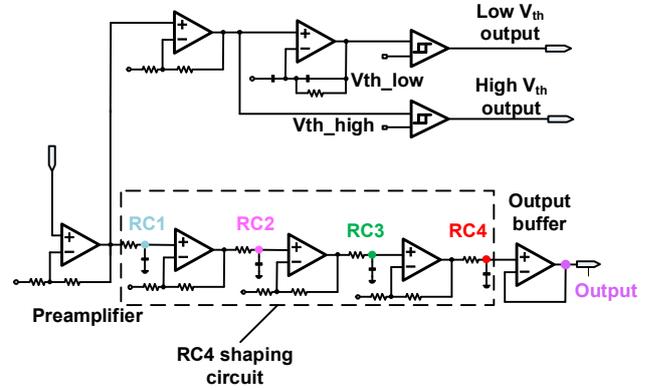

Fig. 5. Structure of the anode channel.

The output signals of the RC$^4$ shaping circuit nodes on cadence spectre platform are presented in Fig. 6. In Fig. 6, the signal changing trend of the RC output nodes can be observed, which meets the design requirements.

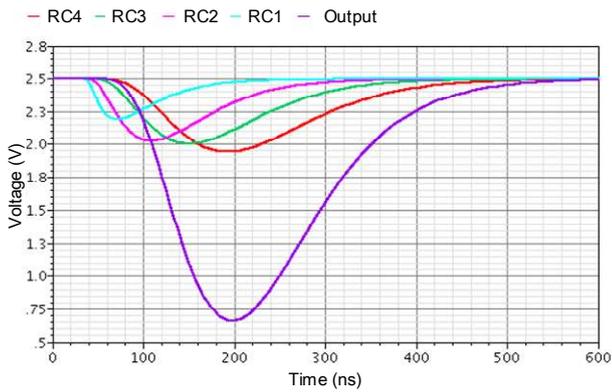

Fig. 6. Output signals of the RC$^4$ shaping circuit nodes on cadence spectre.

## C. Amplification and Shaping ASIC Testing

The block diagram of the test system used to test the amplification and shaping ASIC is presented in Fig. 7. In the test system, the PMT output signals from the signal source are fed to the attenuator to simulate the dynamic range 1-4000 P.E. Then, the signals are fed to the ASIC via the 30-m cable for time and charge measurements. The output time and charge signals are transmitted to the ADC and data transmission board, and then to the computer for further analysis.

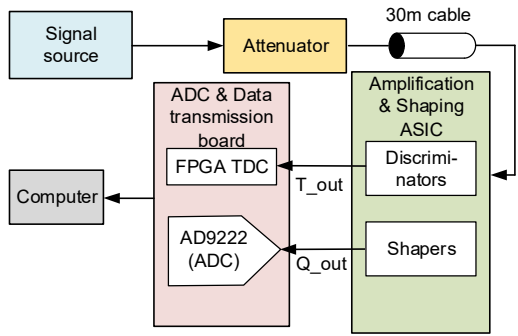

Fig. 7. Block diagram of the amplification and shaping ASIC test system.

As shown in Fig. 7, the output signals from the discriminators are digitized by the FPGA TDC, and the output signals from the shapers are digitized by the AD9222, which is a 12-bit commercial ADC with a sampling rate of 62.5 Mega samples per second.

A total of 12 channels were tested. The results of the charge resolution, time resolution, and time delay of the amplification and shaping ASIC are presented in Fig. 8, Fig. 9, and Fig. 10, respectively. The charge measurement resolution is the ratio of the RMS value of the ADC code to the average value of the ADC code. The line delay method is used to calculate the time resolution because, in that way, the signal source jitter can be eliminated.

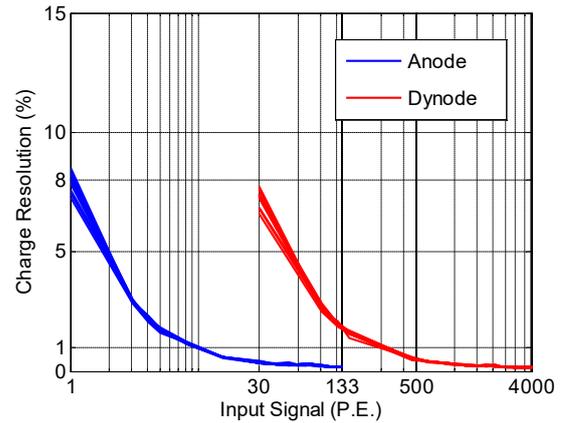

Fig. 8. Charge resolution of the amplification and shaping ASIC.

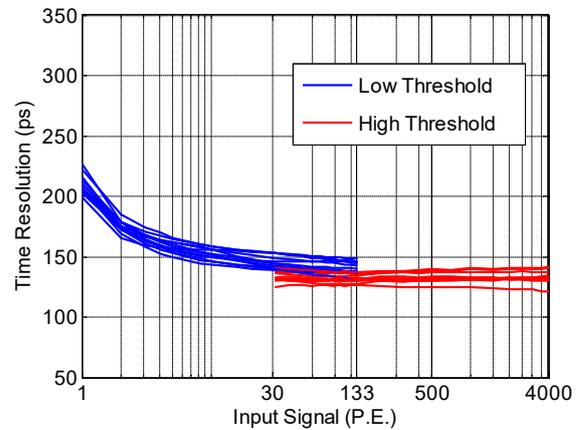

Fig. 9. Time resolution of the amplification and shaping ASIC.

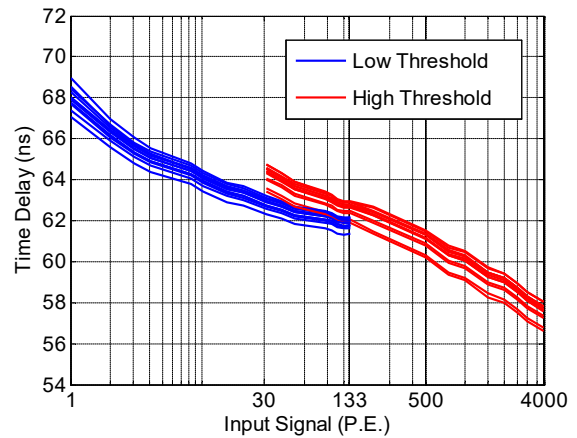

Fig. 10. Time delay of the amplification and shaping ASIC.

Test results indicate that the charge resolution is better than 10% at 1 P.E. and 1% at 4000 P.E., while the time resolution is better than 250 ps using the low threshold and better than 150 ps using the high threshold. Also, time walk is less than 15 ns in the full dynamic range.

## D. Temperature Tests

Additional tests were conducted to evaluate the performance of the ASIC under different temperatures. Test results of the charge resolution, time resolution, and time delay of the amplification and shaping ASIC under different temperatures, which are 0°C, 25°C and 45°C, are shown in Fig. 11, Fig. 12, and Fig. 13, respectively.

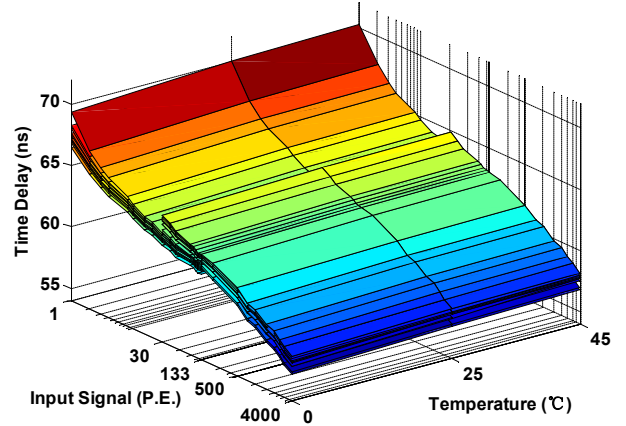

Fig. 13. Time delay of the amplification and shaping ASIC under different temperatures.

In Figs. 11-13, it can be noticed that the amplification and shaping ASIC has a good temperature drift performance. And as presented before, the charge measurement resolution is better than 10% at 1 P.E. and 1% at 4000 P.E. and the time resolution is better than 250 ps while the time walk is less than 15 ns in the full dynamic range under different temperatures.

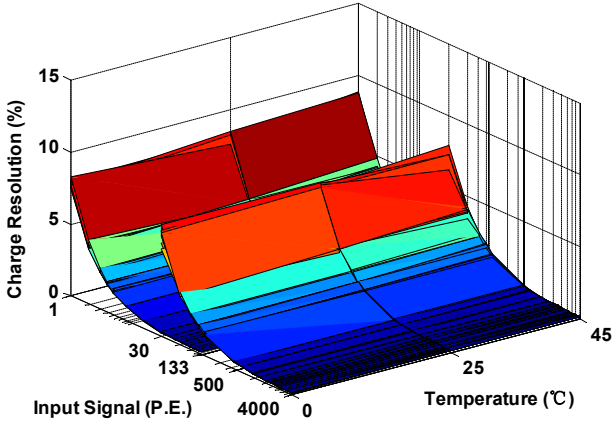

Fig. 11. Charge resolution of the amplification and shaping ASIC under different temperatures.

## IV. SAR ADC ASIC

### A. Structure of SAR ADC ASIC

According to the features of the shaper output signal, as well as the charge resolution requirement, an ADC ASIC was designed with a 12-bit resolution and a sampling rate of about 30 Msps based on the global foundry 1P6M 180-nm CMOS technology. The structure of the designed ADC including the sampling and holding circuits, capacitive DAC, dynamic comparator, asynchronous SAR logic, reference voltage buffer, and clock generator [19-22] is shown in Fig. 14.

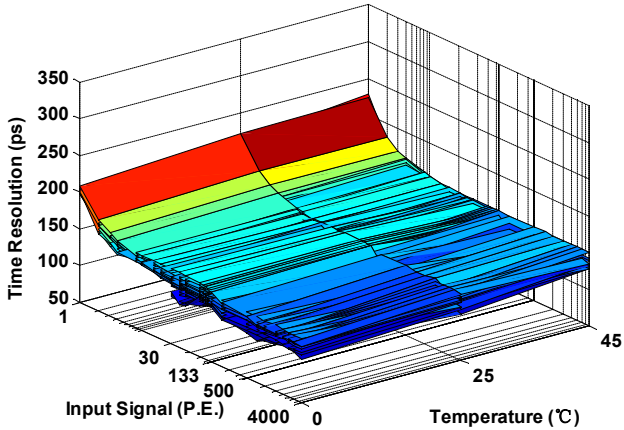

Fig. 12. Time resolution of the amplification and shaping ASIC under different temperatures.

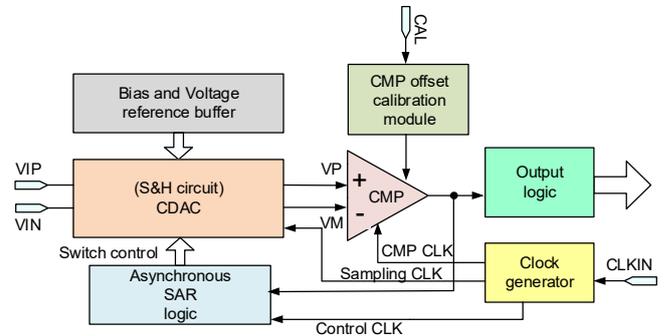

Fig. 14. Structure of the SAR ADC ASIC.

### B. Capacitive DAC

The capacitive DAC of the designed SAR ADC ASIC is presented in Fig. 15. Considering the total number of capacitor and its power consumption, it is more reasonable to adopt two-level weight capacitor DAC structure with integral bridging capacitor.

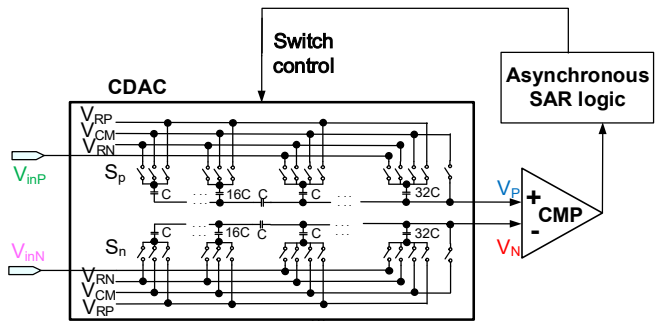

Fig. 15. Structure of the capacitive DAC in the SAR ADC ASIC.

Fig. 16 shows the cadence spectre simulation of the capacitive DAC. CLK is the sampling clock, $V_{inN}$ and $V_{inP}$ are input signals of the CDAC while $V_N$ and $V_P$ are input signals of the comparator.

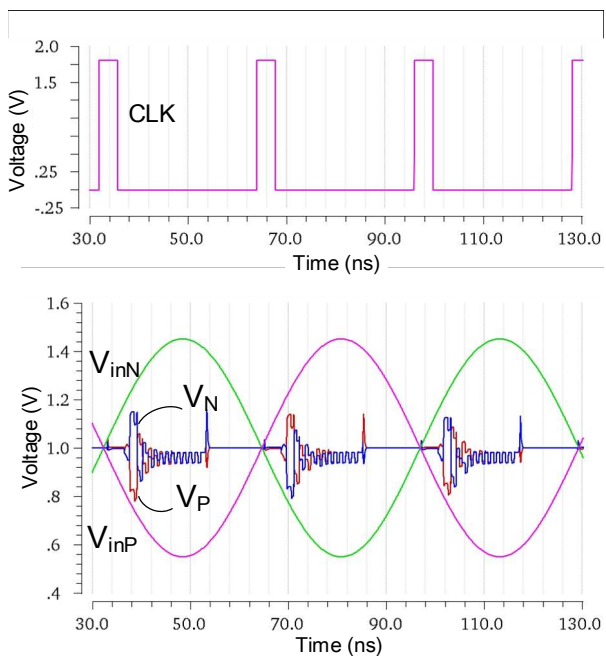

Fig. 16. Cadence spectre simulation of the capacitive DAC.

### C. SAR ADC ASIC Testing

The block diagram of the system for SAR ADC ASIC testing is presented in Fig. 17. The sine waves are generated by the SMA 100A signal source and then fed to the SAR ADC ASIC through the passive LC bandpass filter for digitization. The quantization codes are sent to the data transmission board for data processing and then transferred to the computer via the USB.

The peripheral circuits of the ADC ASIC evaluation board mainly include the input module, output module, clock module, and power module. The Balun (balanced to unbalanced) is used to convert a single-ended output signal of the signal source to a differential signal, and then the voltage of a differential signal is changed by AC coupling to 1 V which is required by the ADC.

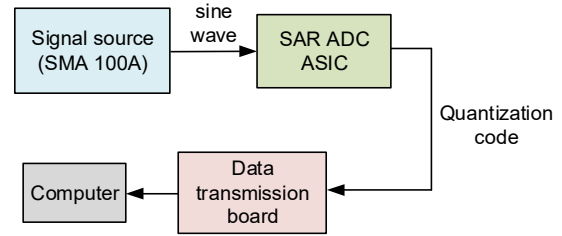

Fig. 17. Block diagram of the system for SAR ADC ASIC testing.

The dynamic performance is tested by the Fast Fourier Transformation method. Test results are shown in Fig. 18, Fig. 19 and Fig. 20. The sampling rate was set to 31.25 Msps. The input signals were sine signals with the frequency of 2.4 MHz, 5 MHz, 8 MHz, 10 MHz, and 15.5 MHz. In order to avoid the coherent sampling, the frequency of an input signal was adjusted slightly.

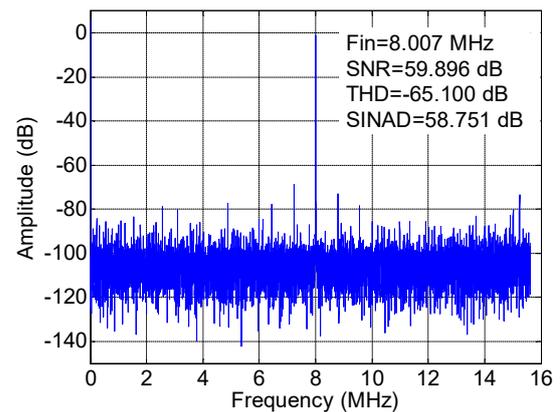

Fig. 18. Typical frequency spectrum at 8.007 MHz.

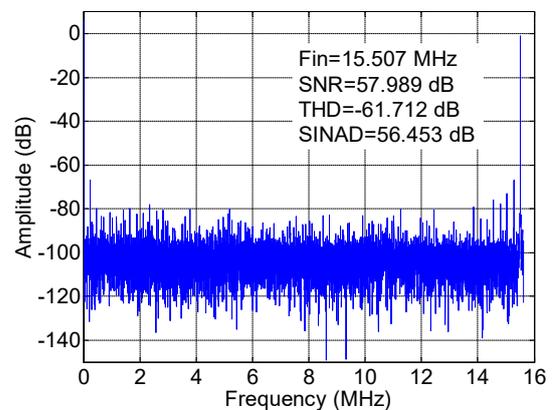

Fig. 19. Typical frequency spectrum at 15.507 MHz.

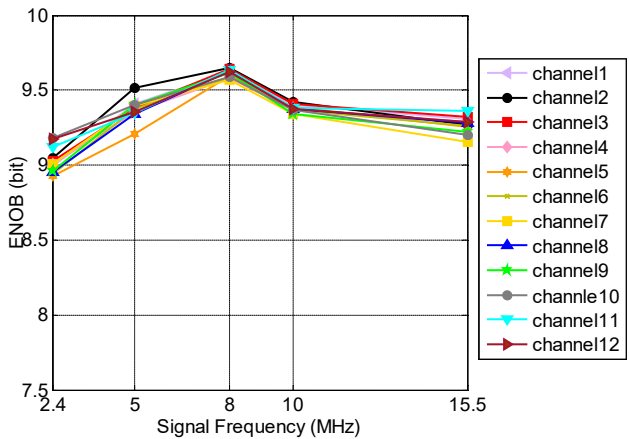

Fig. 20. ENOB dependence on the input signal frequency.

The typical frequency spectrums at the input signal frequencies of 8 MHz and 15.5 MHz are presented in Fig. 18 and Fig. 19, respectively. Fig. 20 shows that the ENOBs of all 12 ADC channels under test are between 8.92 and 9.65 bit.

## V. Combined Test of Two ASICs

The combined tests of two ASIC prototypes were conducted to evaluate the overall charge measurement performance. The test bench used in the tests is presented in Fig. 21.

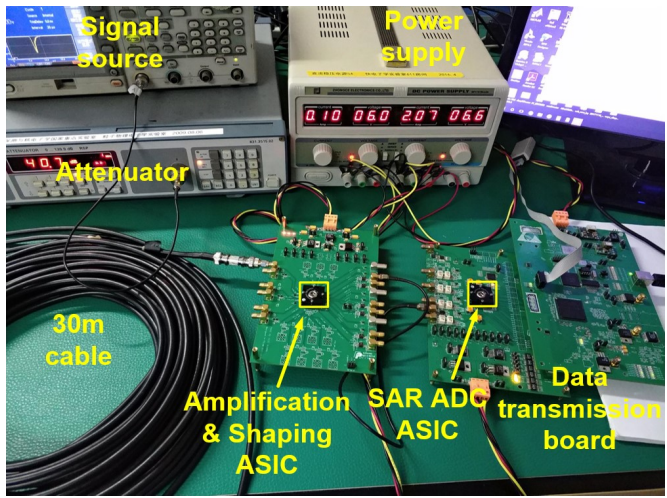

Fig. 21. Test bench of the combined tests.

The sampling rate was set to 31.25 Msps, and the time constant of the amplification and shaping ASIC was set to 40 ns. Only the charge measurement was conducted because the time measurement was accomplished by the discriminator in the amplification and shaping ASIC and FPGA TDC, which was achieved in the data transmission board.

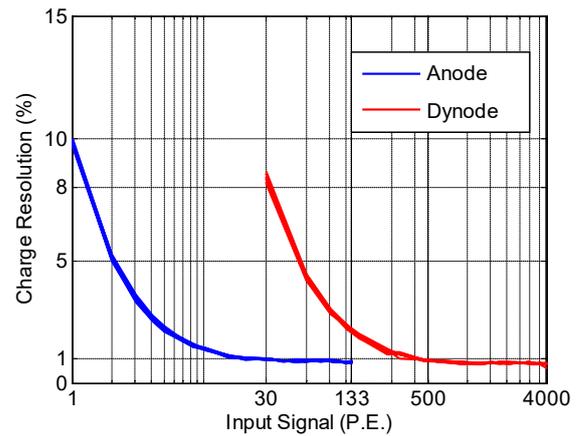

Fig. 22. Charge resolution of the combined tests.

The charge resolution of the tests is presented in Fig. 22, where it can be seen that the charge resolution is better than 10% at 1 P.E. and 1% at 4000 P.E., which meets the requirements of the readout electronics for WCDA.

## VI. Conclusion

Due to the requirements for high precision during the time and charge measurements over a large dynamic range of the LHAASO WCDA, the analysis and design of ASICs are carried out based on the amplification, shaping, and A/D conversion, along with the digital peak detection method. Two prototype ASICs are designed and tested. Tests results indicate that the SAR ADC ASIC can function well at 31.25 MSPS, and its ENOB is between 8.92 bit and 9.65 bit; the time resolution is better than 250 ps RMS, and the overall charge resolution is better than 10% at1 P.E. and 1% at 4000 P.E., which meets the requirements of the readout electronics for LHAASO WCDA.